\documentclass[aps,prl,twocolumn,letterpaper,superscriptaddress,amsmath,amsfonts,showpacs,floatfix,longbibliography]{revtex4-1}
\RequirePackage{graphicx}
\usepackage{placeins}
\usepackage{hyperref}
\usepackage{xcolor}
\newcommand{\fracm}[2]{\mathchoice{\frac{#1}{#2}}{#1/#2}{#1/#2}{#1/#2}}

\begin{document}

\title{Direct verification of the kinetic description of wave turbulence for finite-size systems
dominated by interactions among groups of 6 waves}

\date{\today}

\author{J.\,W.~Banks}
\email{banksj3@rpi.edu}
\affiliation{Mathematics Sciences Department, Rensselaer Polytechnic Institute, 110 8th St., Troy, NY 12180, USA}

\author{T.~Buckmaster}
\email{tjb4@math.princeton.edu}
\affiliation{Department of Mathematics, Princeton University, Princeton, NJ 08544, USA}

\author{A.\,O.~Korotkevich}
\email{alexkor@math.unm.edu}
\affiliation{Department of Mathematics and Statistics, University of New Mexico, MSC01 1115, 1 University of New Mexico, Albuquerque, NM 87131-0001, USA}
\affiliation{L.\,D.~Landau Institute for Theoretical Physics RAS, 2 Kosygin Str., Moscow, 119334, Russian Federation}

\author{G.~Kova\v{c}i\v{c}}
\email{kovacg@rpi.edu}
\affiliation{Mathematics Sciences Department, Rensselaer Polytechnic Institute, 110 8th St., Troy, NY 12180, USA}

\author{J.~Shatah}
\email{shatah@cims.nyu.edu}
\affiliation{Courant Institute of Mathematical Sciences, New York University, 251 Mercer St., New York, NY 10012, USA}


\begin{abstract}\noindent 
The present work considers systems whose dynamics are governed by the nonlinear interactions among groups of 6 nonlinear waves, such as those described by the unforced quintic nonlinear Schr\"odinger equation. Specific parameter regimes in which ensemble-averaged dynamics of such systems with finite size are accurately described by a wave kinetic equation, as used in wave turbulence theory, are theoretically predicted.  In addition, the underlying reasons that the wave kinetic equation may be a poor predictor of wave dynamics outside these regimes are also discussed. These theoretical predictions are directly verified by comparing ensemble averages of solutions to the dynamical equation with corresponding solutions of the wave kinetic equation.
\end{abstract}

\maketitle

The dynamics of large-scale nonlinear systems tend to be so complex that information gleaned from individual trajectories is insufficient to characterize the intrinsic properties of the system. Often such properties are best revealed through statistical measures from  ensembles of trajectories over long times. For particle, plasma, and wave systems, both in and out of equilibrium, kinetic equations have proven to be powerful theoretical tools for ensemble descriptions
\cite{Boltzmann1,Boltzmann2,Nordheim1928,Peierls1929,Vlasov38,Landau46,Hasselmann1962,BS65,Zakharov1967,BN69}.

For weakly nonlinear wave systems, statistical description using the \emph{wave kinetic equation} (WKE) is provided by the \emph{wave turbulence theory} (WTT)~\cite{ZLF1992,Nazarenko2011}, which can be heuristically derived using  perturbation-theoretic arguments~\cite{Hasselmann1962,BS65,Zakharov1967,BN69}.  (In contrast, descriptions of fully developed turbulence~\cite{kolmogorov1941local,Frisch} rely on scaling~\cite{kolmogorov1941local}, models~\cite{gledzer1973system,desnianskii1974simulation,obukhov1974atmos,yamada1988lyapunov,FOIAS2001505}, numerical simulations~\cite{PhysRevLett.100.254504,benzi15}, or are currently unattainable.)  WKEs in WTT have been quite successful in explaining various statistical steady states in systems ranging from surface water waves~\cite{Zakharov1967,Onorato2002,DKZ2003grav,DKZ2004,ZKPR2007,KPRZ2008} to semiconductor lasers~\cite{Lvov1997499}.   

 Applicability of WKEs to dynamically evolving systems has been much less explored (cf. Refs.~\cite{NZ2008,KPZ2019,FV2015}).  
Moreover, strict physical assumptions must be made for WKEs to hold, even in steady state.   These include weak nonlinearity, infinite system size~\cite{ZLF1992,Nazarenko2011}, and  an appropriate moment closure, which is either assumed~\cite{ZLF1992} or follows from assumed validity of multi-scale perturbation expansions~\cite{BN69,BIVEN200398,doi:10.1146/annurev-fluid-122109-160807,doi:10.1142/8269}.   The presence of coherent structures can lead to violations of these assumptions and destroy the validity of the WKE~\cite{MMT1997,CMMT2000,ZVD2001,RB2005}, at times necessitating additional modeling to ensure agreement of the WKE with the underlying physics~\cite{Korotkevich2008PRL,ZKP2009,Korotkevich2012MCS,KPZ2019}.  For dynamically evolving \emph{finite size} systems yet less is known (but see Refs.~\cite{PhysRevE.63.046306,DKZ2003cap,rpi.17522620050101,ZKPD2005,AS2006,KDZ2016,pan_yue_2017}).  Therefore, delineating physical parameter regimes where WKEs accurately describe the dynamics of finite size systems remains a key challenge in WTT.   
 
This challenging problem can be solved either via  ab-initio derivation,  direct ensemble observation (physical or numerical), or a combination of the two.   The first approach was employed in Refs.~\cite{BGHS2021,CoGe2019,DeHa2021a,DeHa2021}, where members of our team and colleagues derived the validity of the WKE for the cubic Nonlinear Schr\"{o}dinger equation on a finite domain in three and more dimensions.
While conceptually important, these results  describe dynamics only up to an infinitesimally small multiple of the characteristic timescale for the WKE. In addition, the work contains no clear delineation of specific regimes for which the WKE holds. 

This letter theoretically delineates regimes of WKE applicability for a \emph{one-dimentional,  finite-size system}, and confirms the resulting predictions on physically meaningful timescales via numerical simulations, thus remedying those deficiencies in prior work.  In the process, \emph{quasi-resonances} are identified as the mechanism underlying the WKE approximation (cf. Refs.~\cite{PhysRevE.63.046306,DKZ2003cap,rpi.17522620050101,ZKPD2005,AS2006,KDZ2016,pan_yue_2017}),  and exact resonances as a mechanism that possibly destroys this approximation.   
Moreover, two sources of coherent structures are described, and their importance is described in terms of the system size $L$: exact resonances, and focusing- or  collapse-like events (cf. Ref.~\cite{doi:10.1007/s00222-021-01067-9}).  None of these properties appear to be true for systems of infinite size.

Specifically, we determine (in)validity regimes of the WKE for one dimensional systems described by the defocusing quintic nonlinear Schr\"{o}dinger equation (DQNLS), 
\begin{equation}
  i u_t + u_{xx} -\mu |u|^4 u = 0,  \label{eq:qNLSE}
\end{equation}
with finite system size modelled by spatial periodicity, $u(x,t)=u(x+L,t)$, with
period $L$. Finite systems subject to other boundary conditions, e.g. Dirichlet or Neumann, are interesting, and the range of applicability of the wave-kinetic theory could be different. This is the subject of future investigation.
%Finite system size is modeled by spatial periodicity, $u(x,t)=u(x+L,t)$, for a given period $L$.
The parameter $\mu>0$ dictates the relative strength of the nonlinearity.  Note that the squared norm $\| u \|^2=\int_0^L |u(x,t)|^2 \,dx$ is a conserved quantity, and due to scaling symmetries of Eq.~\eqref{eq:qNLSE}, we are free to set $\|u\|=L^{\fracm12}$ without loss of generality.  

Due to the nature of the nonlinearity in Eq.~\eqref{eq:qNLSE}, the dynamics of DQNLS waves are dominated by six-wave interactions.   Therefore, despite its simplicity, the considered model has direct relevance to physical systems dominated by six-wave interactions: for example Kelvin waves in superfluid turbulence~\cite{KS2004},  and small fluctuations around both the zero electric field and stable pulses in one-dimensional nonlinear optics~\cite{LBNR2012}; see \footnote{The leading-order terms in the small-amplitude expansion of the governing equations in both Refs.~\cite{KS2004} and~\cite{LBNR2012} yield the dispersion relation $\omega=k^2$ in an appropriate scale range.  For this relation, it is known that no resonant interactions among groups of four waves exist in one spatial dimension.   Therefore, the cubic terms, which are the first nonlinear correction in the governing equation, can be eliminated using a canonical transformation~\cite{ZLF1992}.  The next-order, quintic terms, together with the corresponding resonant interactions among groups of six waves, thus dominate the dynamics. Resonant interactions among groups of eight waves stem from yet higher-order terms and can thus be neglected.}.

The WKE corresponding to Eq.~\eqref{eq:qNLSE} describes the time evolution of the \emph{wave action},  $n_k(t)=\langle \left\vert a_k(t)\right\vert^2\rangle$,  where $a_k(t)$ is the (complex) amplitude  of the wave with wavenumber $k$, and the angle brackets represent averaging over ensembles of initial waves.   Due to this system's finite size, each wavenumber $k$ is an integer multiple of  $\Delta k = 2\pi/L$, and the wave amplitudes are defined via the plane-wave expansion 
\begin{equation}
  u(x,t) = \frac{1}{L^{\fracm 12}}\sum\limits_{k} a_k(t)e^{i(kx-\omega_k t)}, \label{eq:Fourier_def}
\end{equation}
where $\omega_k=k^2$ is the linear dispersion relation for Eq.~\eqref{eq:qNLSE}.   The factor $L^{-\fracm 12}$  in Eq.~\eqref{eq:Fourier_def} is used with an eye on the large $L$ limit, required for the WKE description.   

To showcase a simple, heuristic derivation of WKE starting from our finite-size system, we assume the phases of  wave amplitudes $a_k(t)$  to satisfy the \emph{random phase approximation} (RPA); i.e., in  the second-order perturbation terms we treat phases as independent variables.   (Alternatively, WKE can be obtained using closure, as discussed above.) 
Together with the fact that $\Vert u \Vert =L^{1/2}$, the RPA implies with a high probability \begin{equation}\label{eq:maxu}
  \max_x |u(x,t)| = \mathcal O\left(1\right), \quad \max_kn_k(t) = \mathcal O\left(1\right).
\end{equation}

For waves with finite bandwidth, i.e., those whose wavenumbers $k$ satisfy $| k|\leq  k_{\max}$, the plane-wave expansion in Eq.~\eqref{eq:Fourier_def} and the scaling in Eq.~\eqref{eq:maxu} imply a plausible condition of weak nonlinearity, $\mu\, (\max |u|)^4 = \mu\, \mathcal O(1) \ll \omega_{\max} = k_{\max}^2=\mathcal O(1)$, so
\begin{equation}\label{eq:weaknl}
\mu\ll 1,
\end{equation}
which is also the formal weak-nonlinearity condition in Eq.~\eqref{eq:qNLSE}.
In order to categorize parameter regimes for which we expect  WKE to apply, we link the nonlinearity parameter $\mu$ to the spatial period $L$ via the relation $\mu=L^p$, which is  motivated by the invariance of Eq.~\eqref{eq:qNLSE} to the power-law scaling $x\to \lambda^2 x$, $t\to \lambda^4 t$, $u \to u/\lambda$, with $ \lambda>0$.    The weak nonlinearity condition in Eq.~\eqref{eq:weaknl} thus becomes 
\begin{equation}\label{eq:weak_nonlinearity}
\mu=   L^{p}\ll 1,
\end{equation} implying validity of  WKE for $p<0$.

With scaling $\mu=L^p$ and the RPA, the discrete analog of the WKE describing the evolution of $n_k$ over a time interval $\Delta t\gg 1$ for the finite-size system is~\cite{BGHS2021},
\begin{align}\label{eq:int:step}
  \Delta n_k & = 12L^{2p-4} \sum\limits_{K=0}  \mathfrak{T} \, \frac{\sin^2\left(\Omega \Delta t/2\right)}{\left(\Omega/2\right)^2},
\end{align}
where $\Delta n_k = n_k(\Delta t)-n_k(0)$, and 
\begin{subequations}\label{eq:differences}
\begin{align}
  K & =\sum\limits_{i=0}^2 k_i  -\sum\limits_{i=3}^5k_i; \,\, k_0\equiv k \label{eq:k-difference} \\
  \Omega & = \sum\limits_{i=0}^2\omega_{k_i} -\sum\limits_{i=3}^5\omega_{k_i}, \label{eq:o-difference} \\
\mathfrak{T} & =  \left( \sum\limits_{i=0}^2\frac{1}{n_{k_i}} - \sum\limits_{i=3}^5\frac{1}{n_{k_i}}\right)\prod\limits_{i=0}^5n_{k_i} ,\label{eq:n-difference} 
 \end{align}
\end{subequations}
and $\mathfrak{T}$ is often called the collision term.  [The rather sparse form of $\mathfrak{T}$ in Eq.~\eqref{eq:n-difference} is due to the specific form of the nonlinearity in Eq.~\eqref{eq:qNLSE}, which allows only for the scattering of three waves into three waves and conserves $\| u \|^2$. Scattering of four waves into two waves, or vice versa, present in more general systems dominated by six-wave interactions, is absent from the dynamics of Eq.~\eqref{eq:qNLSE} due to its form of nonlinearity.] To obtain the corresponding WKE for 6 wave interactions, we take the continuum limit of Eq.~\eqref{eq:int:step} by converting the sum to a Riemann sum, and recalling the limit 
\begin{equation}\label{eq:sine-limit}
  \frac{\sin^2(\Omega \Delta t /2)}{(\Omega/2)^2} \to 2\pi \,\Delta t \,\delta(\Omega) \quad \mbox{for} \quad \Delta t \gg 1,
\end{equation}  
where $\delta(\cdot)$ is the Dirac delta.   The WKE is thus
\begin{equation}\label{eq:WKE}
  \frac{d n_k}{d \tau}  =  \int_{-\infty}^\infty \mathfrak{T} \, \delta\left(K\right) \, \delta\left(\Omega\right) \, d\mathbf{k},
\end{equation}
where $d\mathbf{k} = d k_1 \, d k_2\, d k_3 \,d k_4\, d k_5$, $\tau= t/\tau_{kin}$, and $\tau_{kin} =\fracm \pi{6\mu^2}= \fracm\pi{6L^{2p}}$ is the \emph{kinetic time scale} on which the wave actions experience $\mathcal O(1)$ changes.   Note that for the difference $\Delta n_k/\Delta t$ to become the $\tau$-derivative, the inequality $\tau_{kin}\gg \Delta t$ must hold.   This inequality confirms that ensembles of systems described by Eq.~\eqref{eq:qNLSE} evolve slowly, and is also consistent with the small-nonlinearity condition in Eq.~\eqref{eq:weak_nonlinearity}.  Nevertheless, note that Eq.~\eqref{eq:int:step} and the initial RPA only guarantee the validity of WKE in Eq.~\eqref{eq:WKE} on possibly very short $\tau$-scales.  On $\tau$-scales of length $\mathcal  O(1)$, we assume RPA for convenience, or else WKE can \emph{formally} be obtained using the appropriate closure as mentioned above.   However, its validity must be verified by numerical simulations, which we carry out  below. 

Importantly, while the limit in Eq.~\eqref{eq:sine-limit} holds for small $\Omega$, it is incorrect at $\Omega=0$ where the limit is simply $\Delta t^2$. This observation has important consequences for the validity of WKE in Eq.~\eqref{eq:WKE}, and implies that the largest contributions to Eq.~\eqref{eq:WKE} are made by \emph{quasi-resonant} terms in Eq.~\eqref{eq:int:step}, i.e., those terms for which the frequency difference, $\Omega$, is small but does not vanish (cf. Refs.~\cite{PhysRevE.63.046306,DKZ2003cap,rpi.17522620050101,ZKPD2005,AS2006,KDZ2016,pan_yue_2017}).  In fact, the $\Omega$-width of the function described by the ratio on the left-hand side of  Eq.~\eqref{eq:sine-limit} is $1/\Delta t$.   However, because that function has a point of discontinuity at $\Omega=0$, terms corresponding to the \emph{exact resonances}, where both $K$ and $\Omega$ vanish simultaneously, should contribute additional terms of size $\mathcal O\left(\Delta t^2L^{2p-4}\right)$ to Eq.~\eqref{eq:int:step}.  Their inclusion would indicate the possibility of linearly growing terms in Eq.~\eqref{eq:WKE}, whose effect has not been accounted for.  To estimate the cumulative effect of these neglected terms, notice that each term is of size $\mathcal O\left(\Delta t L^{2p-4} \right)$, and a naive count of their number is $\mathcal O\left(L^2\right)$. [A more accurate count, which follows from  number theoretic arguments described in Ref.~\cite{BuGeHaSh18}, is $\mathcal O\left(L^2\ln L\right)$.] Therefore their cumulative contribution will be negligible, and thus the WKE in Eq.~\eqref{eq:WKE} will be valid, provided $\tau_{kin} L^{2p-2}  \ll 1/\tau_{kin}$, i.e., $\tau_{kin} = \mathcal O\left(L^{-2p}\right)\ll L^{1-p}$, and thus $p>-1$.  Importantly, this argument implies that WKE in Eq.~\eqref{eq:WKE} may not be valid for all times, but may  break down at kinetic times $\tau=\mathcal O\left(L^{p+1}\right)$, i.e., $t=\mathcal O\left(L^{1-p}\right) $  in physical units of time. This indicates the validity of the WKE before a breakdown time which increases with $L$.

\begin{figure*}
  \includegraphics[height=1.9in]{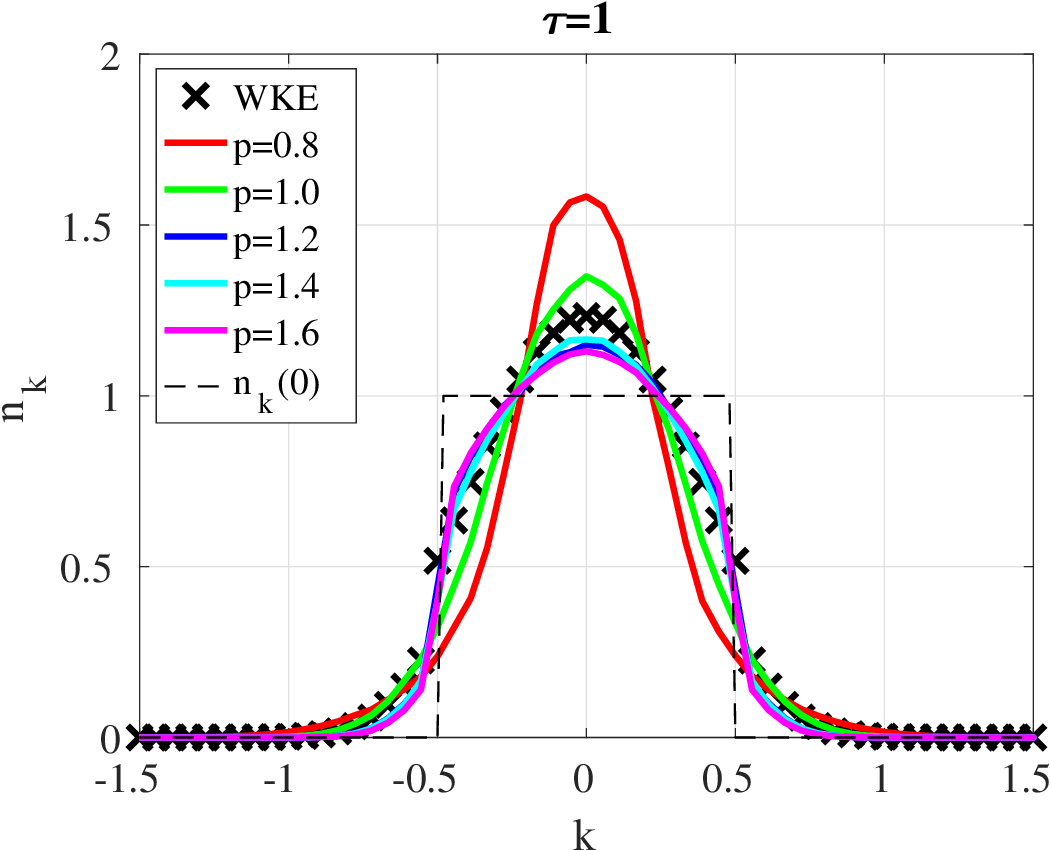}
  \includegraphics[height=1.9in]{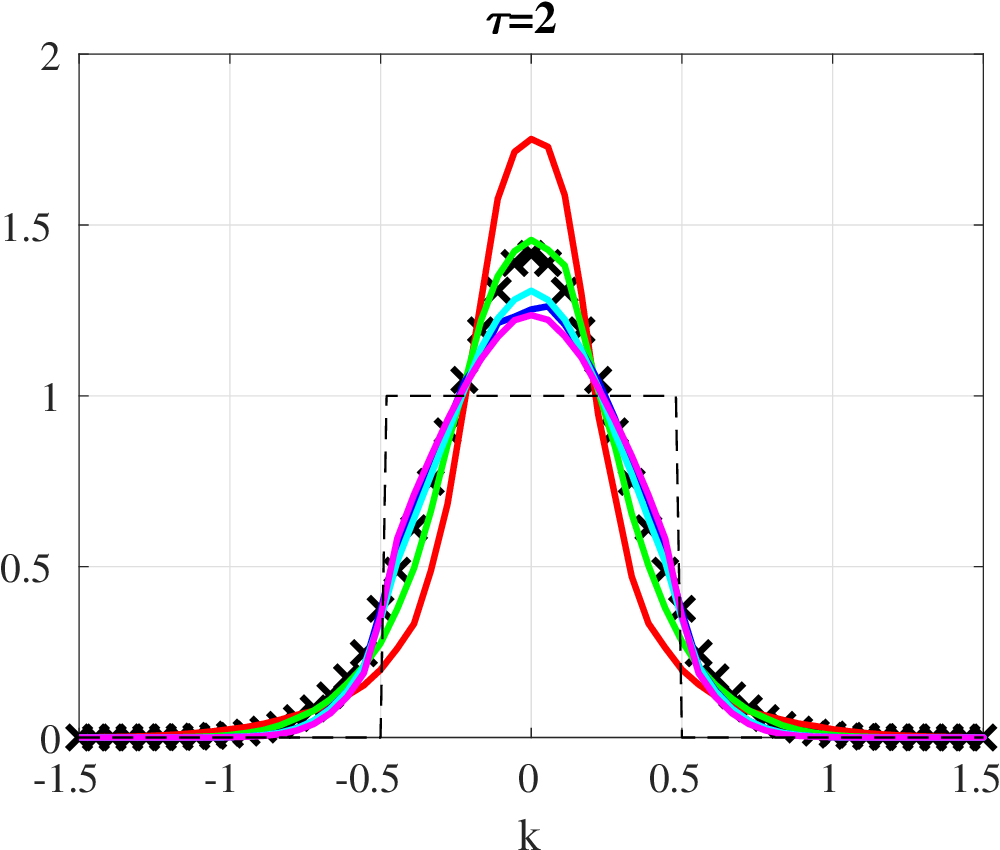}
  \includegraphics[height=1.9in]{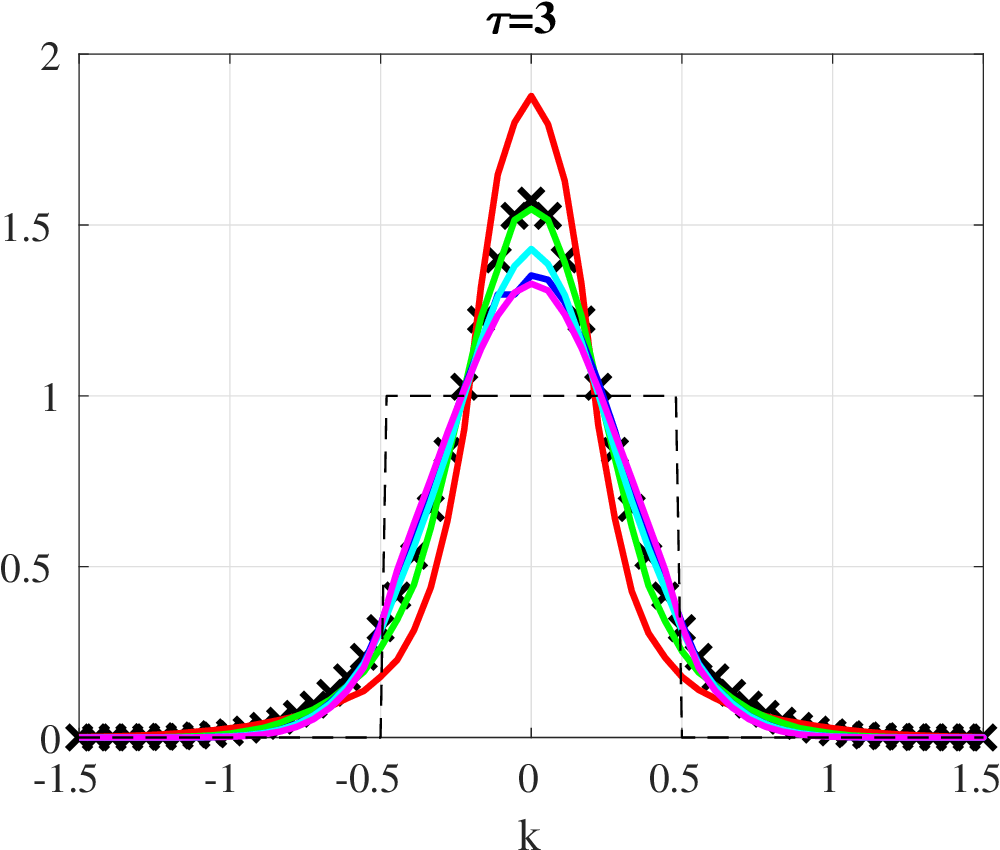}
\caption{\label{fig:DYN_vs_KIN} Comparison of averaged squared amplitudes of harmonics from simulations of DQNLS and WKE for different values of parameter $p$.}
\end{figure*}

\begin{figure}
  \includegraphics[width=3in]{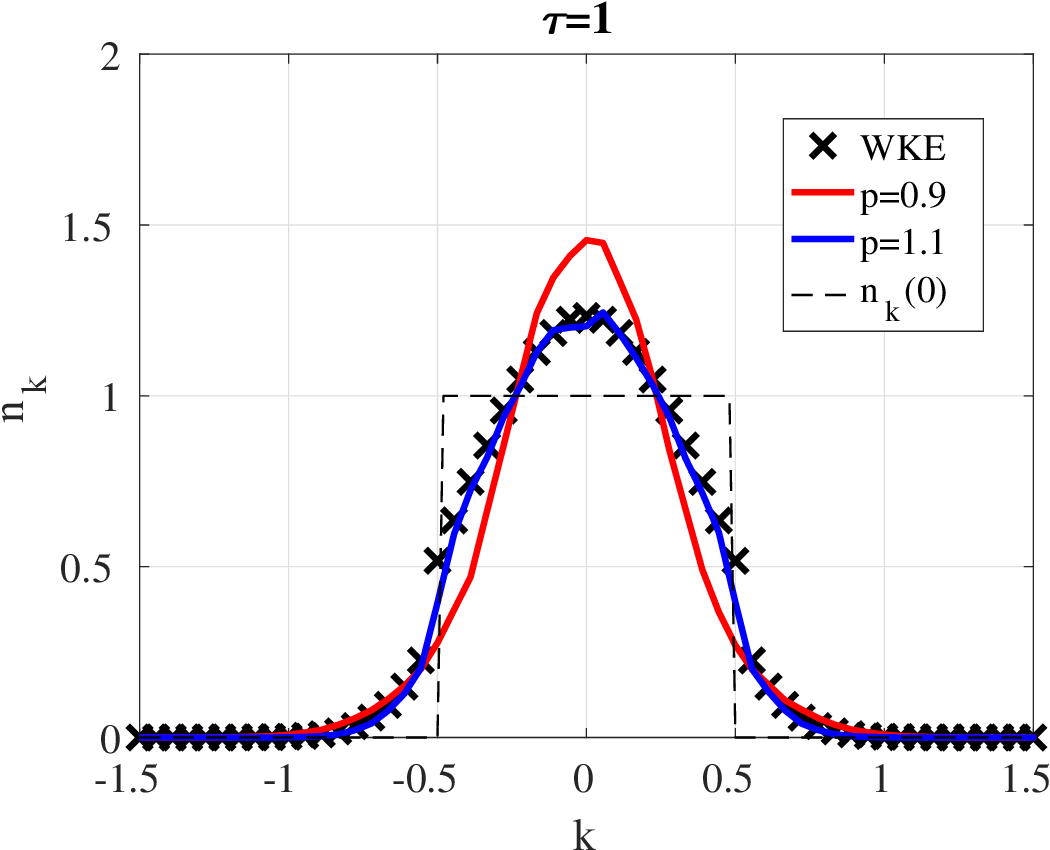}
\caption{\label{fig:DYN_smallp} Solutions of DQNLS at different times for $p=-1.1$, slightly below the threshold in~\eqref{eq:p-condition} and for $p=-0.9$, slightly above the threshold in~\eqref{eq:p-condition}. In both cases $t=\tau_{kin}$ and $L=80$.}
\end{figure}

The temporal bound discussed above also sets a lower bound, $\mu \gg 1/L$, on the strength of the nonlinearity needed for the WKE dynamics to reasonably approximate the ensemble-averaged dynamics of the periodic system with period $L$.   Smaller nonlinearity implies there are insufficient quasi-resonances to generate dynamics describable by WKE. The (perhaps even fewer) exact resonances, however, may instead generate growing observable dynamics [such as those shown in Figs.~\ref{fig:DYN_vs_KIN} and~\ref{fig:DYN_smallp} below with $p=-1.2$ and $p=-1.1$, respectively].   These are known as mesoscopic turbulence~\cite{ZKPD2005}.
The discussion in this and the previous paragraph thus appears to be in contrast with properties of infinite size systems.

We note that a WKE for capillary waves in finite basins that takes into account quasi-resonances via resonant broadening was developed in Ref.~\cite{pan_yue_2017}. 

We now proceed with a numerical determination of the regimes for which the WKE gives a valid description of the dynamics governing ensembles of DQNLS waves on time intervals spanning several kinetic timescales.   We expect WKE to apply to waves emerging from any initial conditions whose plane-wave amplitudes $a_k(0)$ satisfy the RPA. Therefore, as a particularly severe test, we choose discontinuous initial wave amplitudes such that $a_k(0) = C e^{i\gamma_k}$ for the wavenumbers $k$ in some range $| k|\leq  \fracm 12$ and $\gamma_k$ drawn from the uniform distribution of angles on $0\leq\gamma_k<2\pi$, and $a_k(0)=0$ for $|k|> \fracm 12$.  Here, $C>0$ is a constant selected so that $\|u\|=L^{1/2}$.  

\begin{figure}
  \includegraphics[width=3in]{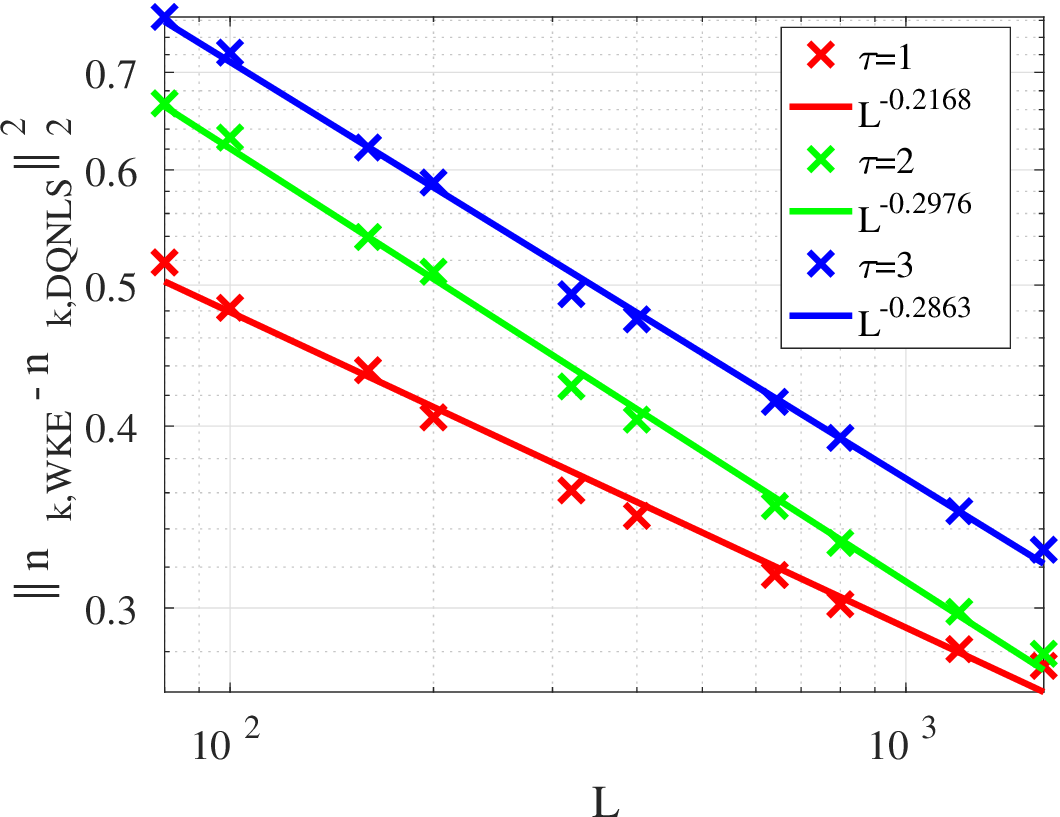}
  \caption{\label{fig:DYN_vs_KIN_error} Mismatch between DQNLS and WKE as a function of $L$ for the case $p=-0.6$ at $\tau=\tau_{kin}$, $2\tau_{kin}$, and $3\tau_{kin}$.}
\end{figure}

Above, we theoretically determined that average ensemble dynamics of DQNLS waves should be well approximated by WKE in Eq.~\eqref{eq:WKE} for $p$ satisfying
\begin{equation}
-1 < p < 0 . \label{eq:p-condition}
\end{equation}
Using the intermediate value $p=-0.6$~\footnote{We note that with  $-1<p<0$ fixed, the requirement $\tau_{kin}\gg 1$ provides an additional practical bound on $L$ for our simulations, which degenerates at $p=0$.},   we proceed to show the correspondence between averaged DQNLS wave ensembles and the corresponding wave actions, $n_{k,\text{WKE}}$ in the limit of large $L$, which is necessary to guarantee that the weak nonlinearity condition in Eq.~\eqref{eq:weak_nonlinearity} is satisfied. Wave ensembles, $n_{k,\text{DQNLS}}$, are computed by averaging squared wave-amplitude moduli $\left\vert a_k(t)\right\vert^2$ from 1000 realizations of the random initial phases $\gamma_k$, followed by evolution via Eq.~\eqref{eq:qNLSE}. These ensemble averages are then compared to  wave actions, $n_{k,\text{WKE}}$, obtained from WKE in Eq.~\eqref{eq:WKE}. The discrepancy between $n_{k,DQNLS}$ and $n_{k,WKE}$ is measured in  the squared norm in wavenumber space [defined as $\Vert \Delta n_k (t)\Vert^2 = \int_{-\infty}^\infty |\Delta n_k(t)|^2 dk$] and the results are presented in Fig.~\ref{fig:DYN_vs_KIN_error}. Also shown in the figure are least squares linear fits to the data on log-log scale. This evidence establishes convergence of wave ensembles derived form the DQNLS in Eq.~\eqref{eq:qNLSE}, and the wave action defined from the WKE in Eq.~\eqref{eq:WKE}, in the limit of large $L$.

Having shown agreement of the WKE and DQNLS for $L\gg 1$, we use the same ensemble averaging process to probe the validity of the WKE for a set of $p$ using finite but large $L$ and for predictions on time intervals of durations $\tau_{kin}$,  $2\tau_{kin}$, and $3\tau_{kin}$.  In addition to values of $p$ where good agreement is expected, we include results for the borderline case $p=-1$, and the case $p=-1.2$ which lies below the WKE validity range given by Eq.~\eqref{eq:p-condition}, i.e., in which exactly resonant interactions  overwhelm the system dynamics before weakly turbulent dynamics describable by WKE in Eq.~\eqref{eq:WKE} could emerge.  Note that for larger values of $p$, it is important to consider large system size $L$ not only to satisfy the weak nonlinearity condition in Eq.~\eqref{eq:weak_nonlinearity}, but also to avoid DQNLS waves that focus sharply towards a possible singularity.   Clearly these focusing waves cannot be included in ensembles exhibiting weakly turbulent behavior.  Fortunately, for fixed $p$, the likelihood of encountering such waves seems to decrease rapidly with $L$, and so by choosing a sufficiently large $L$, no focusing waves are encountered in our ensembles.

Results for all cases are presented in Fig.~\ref{fig:DYN_vs_KIN}, which show that for values of $p$ within the WKE validity range in Eq.~\eqref{eq:p-condition}, the best agreement between the ensemble averaged DQNLS wave dynamics and their description by the WKE in Eq.~\eqref{eq:WKE} occurs at moderate values of the wavenumber $k$, as expected.   At these values, the agreement is almost perfect, and  WKE even captures the remnants of the initial jump in the amplitudes of the individual plane-wave components at $k=\pm1/2$.   For small wavenumbers $k$, only excessively large ensemble sizes would improve the agreement. For very large wavenumbers no quantitative agreement other than vanishing smallness of both wave action measures, $n_{k,DQNLS}$ and $n_{k,WKE}$, is expected or seen. For the borderline case $p=-1$ and the invalid case $p=-1.2$, the ensemble averaged wave system dynamics appear to exhibit an initial tendency towards fast focusing and later growth slowdown, neither of which is captured by WKE.   In particular, the case $p=-1.2$ waves overshoot the WKE prediction for small wavenumbers $k$, and both cases undershoot the WKE predictions in the moderate $k$ regime in which the best agreement is expected.

In order to more closely investigate the transition in the dynamical behavior of DQNLS wave ensembles and the approximating WKE dynamics at the parameter value $p=-1$, we take a closer look at both for  $p=-1.1$ and $p=-0.9$, which are located close to this transition but on the opposite sides of it. Fig.~\ref{fig:DYN_smallp} shows that for the parameter value $p=-0.9$, which lies just over the threshold of the WKE validity range in Eq.~\eqref{eq:p-condition} at $p=-1$,  the WKE gives an excellent approximation to the ensemble averaged shape and dynamics of the corresponding DQNLS waves. In particular, at the time $t=\tau_{kin}$, the WKE captures well the diminishing initial discontinuity of the waves and also accurately approximates the ensemble averaged squared wave amplitudes.   This accuracy continues at the time $t=2\tau_{kin}$. This is not the case for the parameter value  $p=-1.1$, which lies just below the left threshold in Eq.~\eqref{eq:p-condition}.   The ensemble averaged wave system dynamics in this case again appear to exhibit an initial tendency towards fast focusing and later growth slowdown not captured by WKE.  Thus, comparing the behavior of DQNLS wave ensembles and WKE solutions in these two fairly close parameter regimes clearly illustrates the threshold behavior predicted in Eq.~\eqref{eq:p-condition}.  

Simulations of the DQNLS were performed on the periodic domain $x\in[-L/2,L/2]$, with 12th-order accurate central finite differences and 6th-order accurate explicit Runge-Kutta time stepping. With $\Delta t \sim h^2$ chosen for stability, the accuracy of the overall method is $O(h^{12})$. The number of discretization points is chosen to be $N = |\lceil 5L \rceil|+2$, where $|\lceil \cdot \rceil|$ indicates the next larger even integer, which is sufficiently fine that the numerical solutions remain accurate to nearly machine precision throughout the simulation. For a typical realization, the relative errors of the conserved squared norm $\Vert u\Vert^2$ and Hamiltonian $\Vert u_x\Vert^2+(\mu/3)\Vert u^3\Vert^2$ at $3\tau_{kin}$ are $10^{-11}$ and $10^{-8}$, respectively.

The WKE~\eqref{eq:WKE} was solved numerically using an algorithm inspired by the Webb-Resio-Tracy~\cite{Webb1978,TR1982,RP1991} approach to simulation of WKE for gravity waves. In short, the six-dimensional wavenumber space is scanned, and only those mode sextets which satisfy resonant conditions are retained. Details of the algorithm will be published elsewhere. All simulations of WKE were performed using 81 harmonics.
  
To conclude, in the case of the DQNLS on a finite, periodic domain, we provided a clear delineation of parameter regimes for which its corresponding WKE predicted by WTT is expected to be be an accurate approximation of ensemble-averaged system dynamics. These predictions are verified directly via numerical simulation. Furthermore, we laid out straightforward theoretical justification for our parameter regime predictions, and confirmed that quasi-resonances, not exact resonances, appear to be the mechanism responsible for this accurate approximation.   (Cf. Refs.~\cite{PhysRevE.63.046306,DKZ2003cap,rpi.17522620050101,ZKPD2005,AS2006,KDZ2016,pan_yue_2017}.) The influence of exact resonances, in turn, may destroy the validity of WKE.   We also identified focusing-type events as another possible coherent structure that can destroy the validity of WKE approximation.   A study of phase cross-correlations to yet further exclude any possible coherent structures will be presented in a future publication. 
The arguments used in this letter are malleable and may be used to infer parameter regimes of validity for other WKEs predicted by WTT.

The authors are grateful for support from the Simons Collaboration on Wave Turbulence. We also acknowledge NSF support DMS-1900149, DMS-1615859 and DMS-1363013. The simulations presented in this article were performed using the Princeton Research Computing resources at Princeton University and NYU IT High Performance Computing resources.  We would like to thank P. Kramer and Y. Lvov for useful suggestions. 
%\bibliographystyle{apsrev4-2}
%\bibliography{../surfacewaves}

\begin{thebibliography}{60}%
\makeatletter
\providecommand \@ifxundefined [1]{%
 \@ifx{#1\undefined}
}%
\providecommand \@ifnum [1]{%
 \ifnum #1\expandafter \@firstoftwo
 \else \expandafter \@secondoftwo
 \fi
}%
\providecommand \@ifx [1]{%
 \ifx #1\expandafter \@firstoftwo
 \else \expandafter \@secondoftwo
 \fi
}%
\providecommand \natexlab [1]{#1}%
\providecommand \enquote  [1]{``#1''}%
\providecommand \bibnamefont  [1]{#1}%
\providecommand \bibfnamefont [1]{#1}%
\providecommand \citenamefont [1]{#1}%
\providecommand \href@noop [0]{\@secondoftwo}%
\providecommand \href [0]{\begingroup \@sanitize@url \@href}%
\providecommand \@href[1]{\@@startlink{#1}\@@href}%
\providecommand \@@href[1]{\endgroup#1\@@endlink}%
\providecommand \@sanitize@url [0]{\catcode `\\12\catcode `\$12\catcode
  `\&12\catcode `\#12\catcode `\^12\catcode `\_12\catcode `\%12\relax}%
\providecommand \@@startlink[1]{}%
\providecommand \@@endlink[0]{}%
\providecommand \url  [0]{\begingroup\@sanitize@url \@url }%
\providecommand \@url [1]{\endgroup\@href {#1}{\urlprefix }}%
\providecommand \urlprefix  [0]{URL }%
\providecommand \Eprint [0]{\href }%
\providecommand \doibase [0]{https://doi.org/}%
\providecommand \selectlanguage [0]{\@gobble}%
\providecommand \bibinfo  [0]{\@secondoftwo}%
\providecommand \bibfield  [0]{\@secondoftwo}%
\providecommand \translation [1]{[#1]}%
\providecommand \BibitemOpen [0]{}%
\providecommand \bibitemStop [0]{}%
\providecommand \bibitemNoStop [0]{.\EOS\space}%
\providecommand \EOS [0]{\spacefactor3000\relax}%
\providecommand \BibitemShut  [1]{\csname bibitem#1\endcsname}%
\let\auto@bib@innerbib\@empty
%</preamble>
\bibitem [{\citenamefont {Boltzmann}(ge49)}]{Boltzmann1}%
  \BibitemOpen
  \bibfield  {author} {\bibinfo {author} {\bibfnamefont {L.}~\bibnamefont
  {Boltzmann}},\ }\href@noop {} {\bibfield  {journal} {\bibinfo  {journal}
  {Wien. Ber.}\ }\textbf {\bibinfo {volume} {58}},\ \bibinfo {pages} {517}
  (\bibinfo {year} {1868; {\rm reprinted in Boltzmann {\it Abhandlungen}, {\bf
  1}, Barth, Leipzig,1909, page 49}})}\BibitemShut {NoStop}%
\bibitem [{\citenamefont {Boltzmann}(e316)}]{Boltzmann2}%
  \BibitemOpen
  \bibfield  {author} {\bibinfo {author} {\bibfnamefont {L.}~\bibnamefont
  {Boltzmann}},\ }\href@noop {} {\bibfield  {journal} {\bibinfo  {journal}
  {Wien. Ber.}\ }\textbf {\bibinfo {volume} {66}},\ \bibinfo {pages} {275}
  (\bibinfo {year} {1872; {\rm reprinted in Boltzmann {\it Abhandlungen}, {\bf
  1}, Barth, Leipzig,1909, page 316}})}\BibitemShut {NoStop}%
\bibitem [{\citenamefont {Nordheim}(1928)}]{Nordheim1928}%
  \BibitemOpen
  \bibfield  {author} {\bibinfo {author} {\bibfnamefont {L.~W.}\ \bibnamefont
  {Nordheim}},\ }\href {https://doi.org/10.1098/rspa.1928.0126} {\bibfield
  {journal} {\bibinfo  {journal} {Proc. Royal Soc. A}\ }\textbf {\bibinfo
  {volume} {119}},\ \bibinfo {pages} {689} (\bibinfo {year}
  {1928})}\BibitemShut {NoStop}%
\bibitem [{\citenamefont {Peierls}(1929)}]{Peierls1929}%
  \BibitemOpen
  \bibfield  {author} {\bibinfo {author} {\bibfnamefont {R.}~\bibnamefont
  {Peierls}},\ }\href {https://doi.org/10.1002/andp.19293950803} {\bibfield
  {journal} {\bibinfo  {journal} {Annalen der Physik}\ }\textbf {\bibinfo
  {volume} {395}},\ \bibinfo {pages} {1055} (\bibinfo {year}
  {1929})}\BibitemShut {NoStop}%
\bibitem [{\citenamefont {Vlasov}(1938)}]{Vlasov38}%
  \BibitemOpen
  \bibfield  {author} {\bibinfo {author} {\bibfnamefont {A.~A.}\ \bibnamefont
  {Vlasov}},\ }\href@noop {} {\bibfield  {journal} {\bibinfo  {journal} {J.
  Phys. {USSR}}\ ,\ \bibinfo {pages} {291}} (\bibinfo {year}
  {1938})}\BibitemShut {NoStop}%
\bibitem [{\citenamefont {Landau}(1946)}]{Landau46}%
  \BibitemOpen
  \bibfield  {author} {\bibinfo {author} {\bibfnamefont {L.~D.}\ \bibnamefont
  {Landau}},\ }\href@noop {} {\bibfield  {journal} {\bibinfo  {journal} {J.
  Phys. (USSR)}\ }\textbf {\bibinfo {volume} {10}},\ \bibinfo {pages} {25}
  (\bibinfo {year} {1946})}\BibitemShut {NoStop}%
\bibitem [{\citenamefont {Hasselmann}(1962)}]{Hasselmann1962}%
  \BibitemOpen
  \bibfield  {author} {\bibinfo {author} {\bibfnamefont {K.}~\bibnamefont
  {Hasselmann}},\ }\href@noop {} {\bibfield  {journal} {\bibinfo  {journal}
  {J.\,Fluid\,Mech.}\ }\textbf {\bibinfo {volume} {12}},\ \bibinfo {pages}
  {481} (\bibinfo {year} {1962})}\BibitemShut {NoStop}%
\bibitem [{\citenamefont {Benney}\ and\ \citenamefont {Saffman}(1965)}]{BS65}%
  \BibitemOpen
  \bibfield  {author} {\bibinfo {author} {\bibfnamefont {D.}~\bibnamefont
  {Benney}}\ and\ \bibinfo {author} {\bibfnamefont {P.}~\bibnamefont
  {Saffman}},\ }\href@noop {} {\bibfield  {journal} {\bibinfo  {journal}
  {Proceedings of the Royal Society of London, Series A (Mathematical and
  Physical Sciences)}\ }\textbf {\bibinfo {volume} {289}},\ \bibinfo {pages}
  {301} (\bibinfo {year} {1965})}\BibitemShut {NoStop}%
\bibitem [{\citenamefont {Zakharov}(1967)}]{Zakharov1967}%
  \BibitemOpen
  \bibfield  {author} {\bibinfo {author} {\bibfnamefont {V.~E.}\ \bibnamefont
  {Zakharov}},\ }\href@noop {} {\bibfield  {journal} {\bibinfo  {journal}
  {J.\,Exp.\,Teor.\,Phys.}\ }\textbf {\bibinfo {volume} {24}},\ \bibinfo
  {pages} {740} (\bibinfo {year} {1967})}\BibitemShut {NoStop}%
\bibitem [{\citenamefont {Benney}\ and\ \citenamefont {Newell}(1969)}]{BN69}%
  \BibitemOpen
  \bibfield  {author} {\bibinfo {author} {\bibfnamefont {D.~J.}\ \bibnamefont
  {Benney}}\ and\ \bibinfo {author} {\bibfnamefont {A.~C.}\ \bibnamefont
  {Newell}},\ }\href@noop {} {\bibfield  {journal} {\bibinfo  {journal}
  {Studies in Applied Mathematics}\ }\textbf {\bibinfo {volume} {48}},\
  \bibinfo {pages} {29} (\bibinfo {year} {1969})}\BibitemShut {NoStop}%
\bibitem [{\citenamefont {Zakharov}\ \emph {et~al.}(1992)\citenamefont
  {Zakharov}, \citenamefont {Lvov},\ and\ \citenamefont {Falkovich}}]{ZLF1992}%
  \BibitemOpen
  \bibfield  {author} {\bibinfo {author} {\bibfnamefont {V.~E.}\ \bibnamefont
  {Zakharov}}, \bibinfo {author} {\bibfnamefont {V.~S.}\ \bibnamefont {Lvov}},\
  and\ \bibinfo {author} {\bibfnamefont {G.}~\bibnamefont {Falkovich}},\
  }\href@noop {} {\emph {\bibinfo {title} {Kolmogorov Spectra of Turbulence
  I}}}\ (\bibinfo  {publisher} {Springer-Verlag},\ \bibinfo {address}
  {Berlin},\ \bibinfo {year} {1992})\BibitemShut {NoStop}%
\bibitem [{\citenamefont {Nazarenko}(2011)}]{Nazarenko2011}%
  \BibitemOpen
  \bibfield  {author} {\bibinfo {author} {\bibfnamefont {S.}~\bibnamefont
  {Nazarenko}},\ }\href {https://doi.org/10.1007/978-3-642-15942-8} {\emph
  {\bibinfo {title} {Wave Turbulence}}}\ (\bibinfo  {publisher} {Springer
  Berlin Heidelberg},\ \bibinfo {year} {2011})\BibitemShut {NoStop}%
\bibitem [{\citenamefont {Kolmogorov}(1941)}]{kolmogorov1941local}%
  \BibitemOpen
  \bibfield  {author} {\bibinfo {author} {\bibfnamefont {A.~N.}\ \bibnamefont
  {Kolmogorov}},\ }\href@noop {} {\bibfield  {journal} {\bibinfo  {journal}
  {Doklady Akad. Nauk SSSR}\ }\textbf {\bibinfo {volume} {30}},\ \bibinfo
  {pages} {301} (\bibinfo {year} {1941})}\BibitemShut {NoStop}%
\bibitem [{\citenamefont {Frisch}(1995)}]{Frisch}%
  \BibitemOpen
  \bibfield  {author} {\bibinfo {author} {\bibfnamefont {U.}~\bibnamefont
  {Frisch}},\ }\href {https://doi.org/10.1017/CBO9781139170666} {\emph
  {\bibinfo {title} {Turbulence: The Legacy of A. N. Kolmogorov}}}\ (\bibinfo
  {publisher} {Cambridge University Press},\ \bibinfo {year}
  {1995})\BibitemShut {NoStop}%
\bibitem [{\citenamefont {Gledzer}(1973)}]{gledzer1973system}%
  \BibitemOpen
  \bibfield  {author} {\bibinfo {author} {\bibfnamefont {E.}~\bibnamefont
  {Gledzer}},\ }\href@noop {} {\bibfield  {journal} {\bibinfo  {journal}
  {Soviet Physics Doklady}\ }\textbf {\bibinfo {volume} {18}},\ \bibinfo
  {pages} {216} (\bibinfo {year} {1973})}\BibitemShut {NoStop}%
\bibitem [{\citenamefont {Desnianskii}\ and\ \citenamefont
  {Novikov}(1974)}]{desnianskii1974simulation}%
  \BibitemOpen
  \bibfield  {author} {\bibinfo {author} {\bibfnamefont {V.}~\bibnamefont
  {Desnianskii}}\ and\ \bibinfo {author} {\bibfnamefont {E.}~\bibnamefont
  {Novikov}},\ }\href@noop {} {\bibfield  {journal} {\bibinfo  {journal}
  {Journal of Applied Mathematics and Mechanics}\ }\textbf {\bibinfo {volume}
  {38}},\ \bibinfo {pages} {468} (\bibinfo {year} {1974})}\BibitemShut
  {NoStop}%
\bibitem [{\citenamefont {Obukhov}(1974)}]{obukhov1974atmos}%
  \BibitemOpen
  \bibfield  {author} {\bibinfo {author} {\bibfnamefont {A.~M.}\ \bibnamefont
  {Obukhov}},\ }\href@noop {} {\bibfield  {journal} {\bibinfo  {journal}
  {Bulletin of the Academy of Sciences of the USSR, Atmos. Oceanic Phys.}\
  }\textbf {\bibinfo {volume} {10}},\ \bibinfo {pages} {127} (\bibinfo {year}
  {1974})}\BibitemShut {NoStop}%
\bibitem [{\citenamefont {Yamada}\ and\ \citenamefont
  {Ohkitani}(1988)}]{yamada1988lyapunov}%
  \BibitemOpen
  \bibfield  {author} {\bibinfo {author} {\bibfnamefont {M.}~\bibnamefont
  {Yamada}}\ and\ \bibinfo {author} {\bibfnamefont {K.}~\bibnamefont
  {Ohkitani}},\ }\href@noop {} {\bibfield  {journal} {\bibinfo  {journal}
  {Physical review letters}\ }\textbf {\bibinfo {volume} {60}},\ \bibinfo
  {pages} {983} (\bibinfo {year} {1988})}\BibitemShut {NoStop}%
\bibitem [{\citenamefont {Foias}\ \emph {et~al.}(2001)\citenamefont {Foias},
  \citenamefont {Holm},\ and\ \citenamefont {Titi}}]{FOIAS2001505}%
  \BibitemOpen
  \bibfield  {author} {\bibinfo {author} {\bibfnamefont {C.}~\bibnamefont
  {Foias}}, \bibinfo {author} {\bibfnamefont {D.~D.}\ \bibnamefont {Holm}},\
  and\ \bibinfo {author} {\bibfnamefont {E.~S.}\ \bibnamefont {Titi}},\ }\href
  {https://doi.org/https://doi.org/10.1016/S0167-2789(01)00191-9} {\bibfield
  {journal} {\bibinfo  {journal} {Physica D: Nonlinear Phenomena}\ }\textbf
  {\bibinfo {volume} {152-153}},\ \bibinfo {pages} {505} (\bibinfo {year}
  {2001})},\ \bibinfo {note} {advances in Nonlinear Mathematics and Science: A
  Special Issue to Honor Vladimir Zakharov}\BibitemShut {NoStop}%
\bibitem [{\citenamefont {Arn\`eodo}\ \emph {et~al.}(2008)\citenamefont
  {Arn\`eodo}, \citenamefont {Benzi}, \citenamefont {Berg}, \citenamefont
  {Biferale}, \citenamefont {Bodenschatz}, \citenamefont {Busse}, \citenamefont
  {Calzavarini}, \citenamefont {Castaing}, \citenamefont {Cencini},
  \citenamefont {Chevillard}, \citenamefont {Fisher}, \citenamefont {Grauer},
  \citenamefont {Homann}, \citenamefont {Lamb}, \citenamefont {Lanotte},
  \citenamefont {L\'ev\`eque}, \citenamefont {L\"uthi}, \citenamefont {Mann},
  \citenamefont {Mordant}, \citenamefont {M\"uller}, \citenamefont {Ott},
  \citenamefont {Ouellette}, \citenamefont {Pinton}, \citenamefont {Pope},
  \citenamefont {Roux}, \citenamefont {Toschi}, \citenamefont {Xu},\ and\
  \citenamefont {Yeung}}]{PhysRevLett.100.254504}%
  \BibitemOpen
  \bibfield  {author} {\bibinfo {author} {\bibfnamefont {A.}~\bibnamefont
  {Arn\`eodo}}, \bibinfo {author} {\bibfnamefont {R.}~\bibnamefont {Benzi}},
  \bibinfo {author} {\bibfnamefont {J.}~\bibnamefont {Berg}}, \bibinfo {author}
  {\bibfnamefont {L.}~\bibnamefont {Biferale}}, \bibinfo {author}
  {\bibfnamefont {E.}~\bibnamefont {Bodenschatz}}, \bibinfo {author}
  {\bibfnamefont {A.}~\bibnamefont {Busse}}, \bibinfo {author} {\bibfnamefont
  {E.}~\bibnamefont {Calzavarini}}, \bibinfo {author} {\bibfnamefont
  {B.}~\bibnamefont {Castaing}}, \bibinfo {author} {\bibfnamefont
  {M.}~\bibnamefont {Cencini}}, \bibinfo {author} {\bibfnamefont
  {L.}~\bibnamefont {Chevillard}}, \bibinfo {author} {\bibfnamefont {R.~T.}\
  \bibnamefont {Fisher}}, \bibinfo {author} {\bibfnamefont {R.}~\bibnamefont
  {Grauer}}, \bibinfo {author} {\bibfnamefont {H.}~\bibnamefont {Homann}},
  \bibinfo {author} {\bibfnamefont {D.}~\bibnamefont {Lamb}}, \bibinfo {author}
  {\bibfnamefont {A.~S.}\ \bibnamefont {Lanotte}}, \bibinfo {author}
  {\bibfnamefont {E.}~\bibnamefont {L\'ev\`eque}}, \bibinfo {author}
  {\bibfnamefont {B.}~\bibnamefont {L\"uthi}}, \bibinfo {author} {\bibfnamefont
  {J.}~\bibnamefont {Mann}}, \bibinfo {author} {\bibfnamefont {N.}~\bibnamefont
  {Mordant}}, \bibinfo {author} {\bibfnamefont {W.-C.}\ \bibnamefont
  {M\"uller}}, \bibinfo {author} {\bibfnamefont {S.}~\bibnamefont {Ott}},
  \bibinfo {author} {\bibfnamefont {N.~T.}\ \bibnamefont {Ouellette}}, \bibinfo
  {author} {\bibfnamefont {J.-F.}\ \bibnamefont {Pinton}}, \bibinfo {author}
  {\bibfnamefont {S.~B.}\ \bibnamefont {Pope}}, \bibinfo {author}
  {\bibfnamefont {S.~G.}\ \bibnamefont {Roux}}, \bibinfo {author}
  {\bibfnamefont {F.}~\bibnamefont {Toschi}}, \bibinfo {author} {\bibfnamefont
  {H.}~\bibnamefont {Xu}},\ and\ \bibinfo {author} {\bibfnamefont {P.~K.}\
  \bibnamefont {Yeung}} (\bibinfo {collaboration} {International Collaboration
  for Turbulence Research}),\ }\href
  {https://doi.org/10.1103/PhysRevLett.100.254504} {\bibfield  {journal}
  {\bibinfo  {journal} {Phys. Rev. Lett.}\ }\textbf {\bibinfo {volume} {100}},\
  \bibinfo {pages} {254504} (\bibinfo {year} {2008})}\BibitemShut {NoStop}%
\bibitem [{\citenamefont {Benzi}\ and\ \citenamefont
  {Biferale}(2015)}]{benzi15}%
  \BibitemOpen
  \bibfield  {author} {\bibinfo {author} {\bibfnamefont {R.}~\bibnamefont
  {Benzi}}\ and\ \bibinfo {author} {\bibfnamefont {L.}~\bibnamefont
  {Biferale}},\ }\href {https://doi.org/10.1007/s10955-015-1323-9} {\bibfield
  {journal} {\bibinfo  {journal} {Journal of Statistical Physics}\ }\textbf
  {\bibinfo {volume} {161}},\ \bibinfo {pages} {1351} (\bibinfo {year}
  {2015})}\BibitemShut {NoStop}%
\bibitem [{\citenamefont {Onorato}\ \emph {et~al.}(2002)\citenamefont
  {Onorato}, \citenamefont {Osborne}, \citenamefont {Serio}, \citenamefont
  {Resio}, \citenamefont {Pushkarev}, \citenamefont {Zakharov},\ and\
  \citenamefont {Brandini}}]{Onorato2002}%
  \BibitemOpen
  \bibfield  {author} {\bibinfo {author} {\bibfnamefont {M.}~\bibnamefont
  {Onorato}}, \bibinfo {author} {\bibfnamefont {A.~R.}\ \bibnamefont
  {Osborne}}, \bibinfo {author} {\bibfnamefont {M.}~\bibnamefont {Serio}},
  \bibinfo {author} {\bibfnamefont {D.}~\bibnamefont {Resio}}, \bibinfo
  {author} {\bibfnamefont {A.}~\bibnamefont {Pushkarev}}, \bibinfo {author}
  {\bibfnamefont {V.~E.}\ \bibnamefont {Zakharov}},\ and\ \bibinfo {author}
  {\bibfnamefont {C.}~\bibnamefont {Brandini}},\ }\href@noop {} {\bibfield
  {journal} {\bibinfo  {journal} {Phys.\,Rev.\,Lett.}\ }\textbf {\bibinfo
  {volume} {89}},\ \bibinfo {pages} {144501} (\bibinfo {year}
  {2002})}\BibitemShut {NoStop}%
\bibitem [{\citenamefont {Dyachenko}\ \emph
  {et~al.}(2003{\natexlab{a}})\citenamefont {Dyachenko}, \citenamefont
  {Korotkevich},\ and\ \citenamefont {Zakharov}}]{DKZ2003grav}%
  \BibitemOpen
  \bibfield  {author} {\bibinfo {author} {\bibfnamefont {A.~I.}\ \bibnamefont
  {Dyachenko}}, \bibinfo {author} {\bibfnamefont {A.~O.}\ \bibnamefont
  {Korotkevich}},\ and\ \bibinfo {author} {\bibfnamefont {V.~E.}\ \bibnamefont
  {Zakharov}},\ }\href@noop {} {\bibfield  {journal} {\bibinfo  {journal}
  {JETP~Lett.}\ }\textbf {\bibinfo {volume} {77}},\ \bibinfo {pages} {546}
  (\bibinfo {year} {2003}{\natexlab{a}})},\ \Eprint
  {https://arxiv.org/abs/physics/0308101} {physics/0308101} \BibitemShut
  {NoStop}%
\bibitem [{\citenamefont {Dyachenko}\ \emph {et~al.}(2004)\citenamefont
  {Dyachenko}, \citenamefont {Korotkevich},\ and\ \citenamefont
  {Zakharov}}]{DKZ2004}%
  \BibitemOpen
  \bibfield  {author} {\bibinfo {author} {\bibfnamefont {A.~I.}\ \bibnamefont
  {Dyachenko}}, \bibinfo {author} {\bibfnamefont {A.~O.}\ \bibnamefont
  {Korotkevich}},\ and\ \bibinfo {author} {\bibfnamefont {V.~E.}\ \bibnamefont
  {Zakharov}},\ }\href@noop {} {\bibfield  {journal} {\bibinfo  {journal}
  {Phys.\,Rev.\,Lett.}\ }\textbf {\bibinfo {volume} {92}},\ \bibinfo {pages}
  {134501} (\bibinfo {year} {2004})},\ \Eprint
  {https://arxiv.org/abs/physics/0308099} {physics/0308099} \BibitemShut
  {NoStop}%
\bibitem [{\citenamefont {Zakharov}\ \emph {et~al.}(2007)\citenamefont
  {Zakharov}, \citenamefont {Korotkevich}, \citenamefont {Pushkarev},\ and\
  \citenamefont {Resio}}]{ZKPR2007}%
  \BibitemOpen
  \bibfield  {author} {\bibinfo {author} {\bibfnamefont {V.~E.}\ \bibnamefont
  {Zakharov}}, \bibinfo {author} {\bibfnamefont {A.~O.}\ \bibnamefont
  {Korotkevich}}, \bibinfo {author} {\bibfnamefont {A.}~\bibnamefont
  {Pushkarev}},\ and\ \bibinfo {author} {\bibfnamefont {D.}~\bibnamefont
  {Resio}},\ }\href@noop {} {\bibfield  {journal} {\bibinfo  {journal}
  {Phys.\,Rev.\,Lett.}\ }\textbf {\bibinfo {volume} {99}},\ \bibinfo {pages}
  {164501} (\bibinfo {year} {2007})},\ \Eprint
  {https://arxiv.org/abs/0705.2838} {arXiv:0705.2838} \BibitemShut {NoStop}%
\bibitem [{\citenamefont {Korotkevich}\ \emph {et~al.}(2008)\citenamefont
  {Korotkevich}, \citenamefont {Pushkarev}, \citenamefont {Resio},\ and\
  \citenamefont {Zakharov}}]{KPRZ2008}%
  \BibitemOpen
  \bibfield  {author} {\bibinfo {author} {\bibfnamefont {A.~O.}\ \bibnamefont
  {Korotkevich}}, \bibinfo {author} {\bibfnamefont {A.}~\bibnamefont
  {Pushkarev}}, \bibinfo {author} {\bibfnamefont {D.}~\bibnamefont {Resio}},\
  and\ \bibinfo {author} {\bibfnamefont {V.~E.}\ \bibnamefont {Zakharov}},\
  }\href@noop {} {\bibfield  {journal} {\bibinfo  {journal}
  {Eur.\,J.\,Mech.\,B/Fluids}\ }\textbf {\bibinfo {volume} {27}},\ \bibinfo
  {pages} {361} (\bibinfo {year} {2008})},\ \Eprint
  {https://arxiv.org/abs/physics/0702145} {physics/0702145} \BibitemShut
  {NoStop}%
\bibitem [{\citenamefont {Lvov}\ and\ \citenamefont
  {Newell}(1997)}]{Lvov1997499}%
  \BibitemOpen
  \bibfield  {author} {\bibinfo {author} {\bibfnamefont {Y.~V.}\ \bibnamefont
  {Lvov}}\ and\ \bibinfo {author} {\bibfnamefont {A.~C.}\ \bibnamefont
  {Newell}},\ }\href@noop {} {\bibfield  {journal} {\bibinfo  {journal}
  {Physics Letters A}\ }\textbf {\bibinfo {volume} {235}},\ \bibinfo {pages}
  {499 } (\bibinfo {year} {1997})}\BibitemShut {NoStop}%
\bibitem [{\citenamefont {Newell}\ and\ \citenamefont
  {Zakharov}(2008)}]{NZ2008}%
  \BibitemOpen
  \bibfield  {author} {\bibinfo {author} {\bibfnamefont {A.~C.}\ \bibnamefont
  {Newell}}\ and\ \bibinfo {author} {\bibfnamefont {V.~E.}\ \bibnamefont
  {Zakharov}},\ }\href@noop {} {\bibfield  {journal} {\bibinfo  {journal}
  {Phys.\,Lett.\,A}\ }\textbf {\bibinfo {volume} {372}},\ \bibinfo {pages}
  {4230} (\bibinfo {year} {2008})}\BibitemShut {NoStop}%
\bibitem [{\citenamefont {Korotkevich}\ \emph {et~al.}(2019)\citenamefont
  {Korotkevich}, \citenamefont {Prokofiev},\ and\ \citenamefont
  {Zakharov}}]{KPZ2019}%
  \BibitemOpen
  \bibfield  {author} {\bibinfo {author} {\bibfnamefont {A.~O.}\ \bibnamefont
  {Korotkevich}}, \bibinfo {author} {\bibfnamefont {A.~O.}\ \bibnamefont
  {Prokofiev}},\ and\ \bibinfo {author} {\bibfnamefont {V.~E.}\ \bibnamefont
  {Zakharov}},\ }\href {https://doi.org/10.1134/S0021364019050035} {\bibfield
  {journal} {\bibinfo  {journal} {JETP Lett.}\ }\textbf {\bibinfo {volume}
  {109}},\ \bibinfo {pages} {309} (\bibinfo {year} {2019})},\ \Eprint
  {https://arxiv.org/abs/1808.04953} {1808.04953} \BibitemShut {NoStop}%
\bibitem [{\citenamefont {Falkovich}\ and\ \citenamefont
  {Vladimirova}(2015)}]{FV2015}%
  \BibitemOpen
  \bibfield  {author} {\bibinfo {author} {\bibfnamefont {G.}~\bibnamefont
  {Falkovich}}\ and\ \bibinfo {author} {\bibfnamefont {N.}~\bibnamefont
  {Vladimirova}},\ }\href {https://doi.org/10.1103/PhysRevE.91.041201}
  {\bibfield  {journal} {\bibinfo  {journal} {Phys. Rev. E}\ }\textbf {\bibinfo
  {volume} {91}},\ \bibinfo {pages} {041201} (\bibinfo {year}
  {2015})}\BibitemShut {NoStop}%
\bibitem [{\citenamefont {Biven}\ \emph {et~al.}(2003)\citenamefont {Biven},
  \citenamefont {Connaughton},\ and\ \citenamefont {Newell}}]{BIVEN200398}%
  \BibitemOpen
  \bibfield  {author} {\bibinfo {author} {\bibfnamefont {L.}~\bibnamefont
  {Biven}}, \bibinfo {author} {\bibfnamefont {C.}~\bibnamefont {Connaughton}},\
  and\ \bibinfo {author} {\bibfnamefont {A.}~\bibnamefont {Newell}},\ }\href
  {https://doi.org/https://doi.org/10.1016/S0167-2789(03)00215-X} {\bibfield
  {journal} {\bibinfo  {journal} {Physica D: Nonlinear Phenomena}\ }\textbf
  {\bibinfo {volume} {184}},\ \bibinfo {pages} {98} (\bibinfo {year} {2003})},\
  \bibinfo {note} {complexity and Nonlinearity in Physical Systems -- A Special
  Issue to Honor Alan Newell}\BibitemShut {NoStop}%
\bibitem [{\citenamefont {Newell}\ and\ \citenamefont
  {Rumpf}(2011)}]{doi:10.1146/annurev-fluid-122109-160807}%
  \BibitemOpen
  \bibfield  {author} {\bibinfo {author} {\bibfnamefont {A.~C.}\ \bibnamefont
  {Newell}}\ and\ \bibinfo {author} {\bibfnamefont {B.}~\bibnamefont {Rumpf}},\
  }\href {https://doi.org/10.1146/annurev-fluid-122109-160807} {\bibfield
  {journal} {\bibinfo  {journal} {Annual Review of Fluid Mechanics}\ }\textbf
  {\bibinfo {volume} {43}},\ \bibinfo {pages} {59} (\bibinfo {year} {2011})},\
  \Eprint
  {https://arxiv.org/abs/https://doi.org/10.1146/annurev-fluid-122109-160807}
  {https://doi.org/10.1146/annurev-fluid-122109-160807} \BibitemShut {NoStop}%
\bibitem [{\citenamefont {Shrira}\ and\ \citenamefont
  {Nazarenko}(2013)}]{doi:10.1142/8269}%
  \BibitemOpen
  \bibfield  {author} {\bibinfo {author} {\bibfnamefont {V.}~\bibnamefont
  {Shrira}}\ and\ \bibinfo {author} {\bibfnamefont {S.}~\bibnamefont
  {Nazarenko}},\ }\href {https://doi.org/10.1142/8269} {\emph {\bibinfo {title}
  {Advances in Wave Turbulence}}}\ (\bibinfo  {publisher} {WORLD SCIENTIFIC},\
  \bibinfo {year} {2013})\ \Eprint
  {https://arxiv.org/abs/https://www.worldscientific.com/doi/pdf/10.1142/8269}
  {https://www.worldscientific.com/doi/pdf/10.1142/8269} \BibitemShut {NoStop}%
\bibitem [{\citenamefont {Majda}\ \emph {et~al.}(1997)\citenamefont {Majda},
  \citenamefont {McLaughlin},\ and\ \citenamefont {Tabak}}]{MMT1997}%
  \BibitemOpen
  \bibfield  {author} {\bibinfo {author} {\bibfnamefont {A.~J.}\ \bibnamefont
  {Majda}}, \bibinfo {author} {\bibfnamefont {D.~W.}\ \bibnamefont
  {McLaughlin}},\ and\ \bibinfo {author} {\bibfnamefont {E.~G.}\ \bibnamefont
  {Tabak}},\ }\href {https://doi.org/10.1007/BF02679124} {\bibfield  {journal}
  {\bibinfo  {journal} {J. Nonlin. Sc.}\ }\textbf {\bibinfo {volume} {7}},\
  \bibinfo {pages} {9} (\bibinfo {year} {1997})}\BibitemShut {NoStop}%
\bibitem [{\citenamefont {Cai}\ \emph {et~al.}(2000)\citenamefont {Cai},
  \citenamefont {Majda}, \citenamefont {McLaughlin},\ and\ \citenamefont
  {Tabak}}]{CMMT2000}%
  \BibitemOpen
  \bibfield  {author} {\bibinfo {author} {\bibfnamefont {D.}~\bibnamefont
  {Cai}}, \bibinfo {author} {\bibfnamefont {A.}~\bibnamefont {Majda}}, \bibinfo
  {author} {\bibfnamefont {D.}~\bibnamefont {McLaughlin}},\ and\ \bibinfo
  {author} {\bibfnamefont {E.}~\bibnamefont {Tabak}},\ }\href
  {https://doi.org/10.1073/pnas.96.25.14216} {\bibfield  {journal} {\bibinfo
  {journal} {PNAS}\ }\textbf {\bibinfo {volume} {96}},\ \bibinfo {pages}
  {14216} (\bibinfo {year} {2000})}\BibitemShut {NoStop}%
\bibitem [{\citenamefont {Zakharov}\ \emph {et~al.}(2001)\citenamefont
  {Zakharov}, \citenamefont {Vasiliev},\ and\ \citenamefont
  {Dyachenko}}]{ZVD2001}%
  \BibitemOpen
  \bibfield  {author} {\bibinfo {author} {\bibfnamefont {V.~E.}\ \bibnamefont
  {Zakharov}}, \bibinfo {author} {\bibfnamefont {O.~A.}\ \bibnamefont
  {Vasiliev}},\ and\ \bibinfo {author} {\bibfnamefont {A.~I.}\ \bibnamefont
  {Dyachenko}},\ }\href {https://doi.org/10.1134/1.1358420} {\bibfield
  {journal} {\bibinfo  {journal} {JETP\,Lett.}\ }\textbf {\bibinfo {volume}
  {73}},\ \bibinfo {pages} {63} (\bibinfo {year} {2001})}\BibitemShut {NoStop}%
\bibitem [{\citenamefont {Rumpf}\ and\ \citenamefont {Biven}(2005)}]{RB2005}%
  \BibitemOpen
  \bibfield  {author} {\bibinfo {author} {\bibfnamefont {B.}~\bibnamefont
  {Rumpf}}\ and\ \bibinfo {author} {\bibfnamefont {L.}~\bibnamefont {Biven}},\
  }\href {https://doi.org/10.1016/j.physd.2005.04.012} {\bibfield  {journal}
  {\bibinfo  {journal} {Physica D: Nonlinear Phenomena}\ }\textbf {\bibinfo
  {volume} {204}},\ \bibinfo {pages} {188} (\bibinfo {year}
  {2005})}\BibitemShut {NoStop}%
\bibitem [{\citenamefont {Korotkevich}(2008)}]{Korotkevich2008PRL}%
  \BibitemOpen
  \bibfield  {author} {\bibinfo {author} {\bibfnamefont {A.~O.}\ \bibnamefont
  {Korotkevich}},\ }\href@noop {} {\bibfield  {journal} {\bibinfo  {journal}
  {Phys.\,Rev.\,Lett.}\ }\textbf {\bibinfo {volume} {101}},\ \bibinfo {pages}
  {074504} (\bibinfo {year} {2008})},\ \Eprint
  {https://arxiv.org/abs/0805.0445} {arXiv:0805.0445} \BibitemShut {NoStop}%
\bibitem [{\citenamefont {Zakharov}\ \emph {et~al.}(2009)\citenamefont
  {Zakharov}, \citenamefont {Korotkevich},\ and\ \citenamefont
  {Prokofiev}}]{ZKP2009}%
  \BibitemOpen
  \bibfield  {author} {\bibinfo {author} {\bibfnamefont {V.~E.}\ \bibnamefont
  {Zakharov}}, \bibinfo {author} {\bibfnamefont {A.~O.}\ \bibnamefont
  {Korotkevich}},\ and\ \bibinfo {author} {\bibfnamefont {A.~O.}\ \bibnamefont
  {Prokofiev}},\ }\href@noop {} {\bibfield  {journal} {\bibinfo  {journal} {AIP
  Proceedings, CP1168}\ }\textbf {\bibinfo {volume} {2}},\ \bibinfo {pages}
  {1229} (\bibinfo {year} {2009})}\BibitemShut {NoStop}%
\bibitem [{\citenamefont {Korotkevich}(2012)}]{Korotkevich2012MCS}%
  \BibitemOpen
  \bibfield  {author} {\bibinfo {author} {\bibfnamefont {A.~O.}\ \bibnamefont
  {Korotkevich}},\ }\href@noop {} {\bibfield  {journal} {\bibinfo  {journal}
  {Math.\,Comput.\,Simul.}\ }\textbf {\bibinfo {volume} {82}},\ \bibinfo
  {pages} {1228} (\bibinfo {year} {2012})},\ \Eprint
  {https://arxiv.org/abs/0911.0741} {arXiv:0911.0741} \BibitemShut {NoStop}%
\bibitem [{\citenamefont {Connaughton}\ \emph {et~al.}(2001)\citenamefont
  {Connaughton}, \citenamefont {Nazarenko},\ and\ \citenamefont
  {Pushkarev}}]{PhysRevE.63.046306}%
  \BibitemOpen
  \bibfield  {author} {\bibinfo {author} {\bibfnamefont {C.}~\bibnamefont
  {Connaughton}}, \bibinfo {author} {\bibfnamefont {S.}~\bibnamefont
  {Nazarenko}},\ and\ \bibinfo {author} {\bibfnamefont {A.}~\bibnamefont
  {Pushkarev}},\ }\href {https://doi.org/10.1103/PhysRevE.63.046306} {\bibfield
   {journal} {\bibinfo  {journal} {Phys. Rev. E}\ }\textbf {\bibinfo {volume}
  {63}},\ \bibinfo {pages} {046306} (\bibinfo {year} {2001})}\BibitemShut
  {NoStop}%
\bibitem [{\citenamefont {Dyachenko}\ \emph
  {et~al.}(2003{\natexlab{b}})\citenamefont {Dyachenko}, \citenamefont
  {Korotkevich},\ and\ \citenamefont {Zakharov}}]{DKZ2003cap}%
  \BibitemOpen
  \bibfield  {author} {\bibinfo {author} {\bibfnamefont {A.~I.}\ \bibnamefont
  {Dyachenko}}, \bibinfo {author} {\bibfnamefont {A.~O.}\ \bibnamefont
  {Korotkevich}},\ and\ \bibinfo {author} {\bibfnamefont {V.~E.}\ \bibnamefont
  {Zakharov}},\ }\href@noop {} {\bibfield  {journal} {\bibinfo  {journal}
  {JETP~Lett.}\ }\textbf {\bibinfo {volume} {77}},\ \bibinfo {pages} {477}
  (\bibinfo {year} {2003}{\natexlab{b}})},\ \Eprint
  {https://arxiv.org/abs/physics/0308100} {physics/0308100} \BibitemShut
  {NoStop}%
\bibitem [{\citenamefont {Pokorni}(2005)}]{rpi.17522620050101}%
  \BibitemOpen
  \bibfield  {author} {\bibinfo {author} {\bibfnamefont {B.}~\bibnamefont
  {Pokorni}},\ }\emph {\bibinfo {title} {On the dynamics and intermittency in
  the numerical model of gravity surface water waves.}},\ \href
  {https://search.ebscohost.com/login.aspx?direct=true&db=ir00332a&AN=rpi.175226&site=eds-live&scope=site&authtype=sso&custid=s9001156}
  {Ph.D. thesis},\ \bibinfo  {school} {Rensselaer Polytechnic Institute, Troy,
  NY} (\bibinfo {year} {2005})\BibitemShut {NoStop}%
\bibitem [{\citenamefont {Zakharov}\ \emph {et~al.}(2005)\citenamefont
  {Zakharov}, \citenamefont {Korotkevich}, \citenamefont {Pushkarev},\ and\
  \citenamefont {Dyachenko}}]{ZKPD2005}%
  \BibitemOpen
  \bibfield  {author} {\bibinfo {author} {\bibfnamefont {V.~E.}\ \bibnamefont
  {Zakharov}}, \bibinfo {author} {\bibfnamefont {A.~O.}\ \bibnamefont
  {Korotkevich}}, \bibinfo {author} {\bibfnamefont {A.}~\bibnamefont
  {Pushkarev}},\ and\ \bibinfo {author} {\bibfnamefont {A.~I.}\ \bibnamefont
  {Dyachenko}},\ }\href@noop {} {\bibfield  {journal} {\bibinfo  {journal}
  {JETP\,Lett.}\ }\textbf {\bibinfo {volume} {82}},\ \bibinfo {pages} {487}
  (\bibinfo {year} {2005})},\ \Eprint {https://arxiv.org/abs/physics/0508155}
  {physics/0508155} \BibitemShut {NoStop}%
\bibitem [{\citenamefont {Annenkov}\ and\ \citenamefont
  {Shrira}(2006)}]{AS2006}%
  \BibitemOpen
  \bibfield  {author} {\bibinfo {author} {\bibfnamefont {S.~Y.}\ \bibnamefont
  {Annenkov}}\ and\ \bibinfo {author} {\bibfnamefont {V.~I.}\ \bibnamefont
  {Shrira}},\ }\href@noop {} {\bibfield  {journal} {\bibinfo  {journal}
  {Phys.\,Rev.\,Lett.}\ }\textbf {\bibinfo {volume} {96}},\ \bibinfo {pages}
  {204501} (\bibinfo {year} {2006})}\BibitemShut {NoStop}%
\bibitem [{\citenamefont {Korotkevich}\ \emph {et~al.}(2016)\citenamefont
  {Korotkevich}, \citenamefont {Dyachenko},\ and\ \citenamefont
  {Zakharov}}]{KDZ2016}%
  \BibitemOpen
  \bibfield  {author} {\bibinfo {author} {\bibfnamefont {A.~O.}\ \bibnamefont
  {Korotkevich}}, \bibinfo {author} {\bibfnamefont {A.~I.}\ \bibnamefont
  {Dyachenko}},\ and\ \bibinfo {author} {\bibfnamefont {V.~E.}\ \bibnamefont
  {Zakharov}},\ }\href@noop {} {\bibfield  {journal} {\bibinfo  {journal}
  {Physica~D}\ }\textbf {\bibinfo {volume} {321-322}},\ \bibinfo {pages} {51}
  (\bibinfo {year} {2016})},\ \Eprint {https://arxiv.org/abs/1212.2225}
  {1212.2225} \BibitemShut {NoStop}%
\bibitem [{\citenamefont {Pan}\ and\ \citenamefont {Yue}(2017)}]{pan_yue_2017}%
  \BibitemOpen
  \bibfield  {author} {\bibinfo {author} {\bibfnamefont {Y.}~\bibnamefont
  {Pan}}\ and\ \bibinfo {author} {\bibfnamefont {D.~K.~P.}\ \bibnamefont
  {Yue}},\ }\href {https://doi.org/10.1017/jfm.2017.106} {\bibfield  {journal}
  {\bibinfo  {journal} {Journal of Fluid Mechanics}\ }\textbf {\bibinfo
  {volume} {816}},\ \bibinfo {pages} {R1} (\bibinfo {year} {2017})}\BibitemShut
  {NoStop}%
\bibitem [{\citenamefont {{Buckmaster}}\ \emph {et~al.}(pear)\citenamefont
  {{Buckmaster}}, \citenamefont {{Germain}}, \citenamefont {{Hani}},\ and\
  \citenamefont {{Shatah}}}]{BGHS2021}%
  \BibitemOpen
  \bibfield  {author} {\bibinfo {author} {\bibfnamefont {T.}~\bibnamefont
  {{Buckmaster}}}, \bibinfo {author} {\bibfnamefont {P.}~\bibnamefont
  {{Germain}}}, \bibinfo {author} {\bibfnamefont {Z.}~\bibnamefont {{Hani}}},\
  and\ \bibinfo {author} {\bibfnamefont {J.}~\bibnamefont {{Shatah}}},\
  }\href@noop {} {\bibfield  {journal} {\bibinfo  {journal} {Inventiones}\ }
  (\bibinfo {year} {to appear})}\BibitemShut {NoStop}%
\bibitem [{\citenamefont {{Collot}}\ and\ \citenamefont
  {{Germain}}(2019)}]{CoGe2019}%
  \BibitemOpen
  \bibfield  {author} {\bibinfo {author} {\bibfnamefont {C.}~\bibnamefont
  {{Collot}}}\ and\ \bibinfo {author} {\bibfnamefont {P.}~\bibnamefont
  {{Germain}}},\ }\href@noop {} {\bibfield  {journal} {\bibinfo  {journal}
  {arXiv e-prints}\ ,\ \bibinfo {eid} {arXiv:1912.10368}} (\bibinfo {year}
  {2019})},\ \Eprint {https://arxiv.org/abs/1912.10368} {arXiv:1912.10368
  [math.AP]} \BibitemShut {NoStop}%
\bibitem [{\citenamefont {Deng}\ and\ \citenamefont {Hani}(2021)}]{DeHa2021a}%
  \BibitemOpen
  \bibfield  {author} {\bibinfo {author} {\bibfnamefont {Y.}~\bibnamefont
  {Deng}}\ and\ \bibinfo {author} {\bibfnamefont {Z.}~\bibnamefont {Hani}},\
  }\href {https://doi.org/10.1017/fmp.2021.6} {\bibfield  {journal} {\bibinfo
  {journal} {Forum of Mathematics, Pi}\ }\textbf {\bibinfo {volume} {9}},\
  \bibinfo {pages} {e6} (\bibinfo {year} {2021})}\BibitemShut {NoStop}%
\bibitem [{\citenamefont {{Deng}}\ and\ \citenamefont
  {{Hani}}(2021)}]{DeHa2021}%
  \BibitemOpen
  \bibfield  {author} {\bibinfo {author} {\bibfnamefont {Y.}~\bibnamefont
  {{Deng}}}\ and\ \bibinfo {author} {\bibfnamefont {Z.}~\bibnamefont
  {{Hani}}},\ }\href@noop {} {\bibfield  {journal} {\bibinfo  {journal} {arXiv
  e-prints}\ ,\ \bibinfo {eid} {arXiv:2104.11204}} (\bibinfo {year} {2021})},\
  \Eprint {https://arxiv.org/abs/2104.11204} {arXiv:2104.11204 [math.AP]}
  \BibitemShut {NoStop}%
\bibitem [{\citenamefont {Merle}\ \emph {et~al.}(2021)\citenamefont {Merle},
  \citenamefont {Rapha\"{e}l}, \citenamefont {Rodnianski},\ and\ \citenamefont
  {Szeftel}}]{doi:10.1007/s00222-021-01067-9}%
  \BibitemOpen
  \bibfield  {author} {\bibinfo {author} {\bibfnamefont {F.}~\bibnamefont
  {Merle}}, \bibinfo {author} {\bibfnamefont {P.}~\bibnamefont {Rapha\"{e}l}},
  \bibinfo {author} {\bibfnamefont {I.}~\bibnamefont {Rodnianski}},\ and\
  \bibinfo {author} {\bibfnamefont {J.}~\bibnamefont {Szeftel}},\ }\bibfield
  {journal} {\bibinfo  {journal} {Inventiones mathematicae}\ }\href
  {https://doi.org/10.1007/s00222-021-01067-9} {10.1007/s00222-021-01067-9}
  (\bibinfo {year} {2021}),\ \Eprint
  {https://arxiv.org/abs/https://doi.org/10.1007/s00222-021-01067-9}
  {https://doi.org/10.1007/s00222-021-01067-9} \BibitemShut {NoStop}%
\bibitem [{\citenamefont {Kozik}\ and\ \citenamefont
  {Svistunov}(2004)}]{KS2004}%
  \BibitemOpen
  \bibfield  {author} {\bibinfo {author} {\bibfnamefont {E.}~\bibnamefont
  {Kozik}}\ and\ \bibinfo {author} {\bibfnamefont {B.}~\bibnamefont
  {Svistunov}},\ }\href {https://doi.org/10.1103/PhysRevLett.92.035301}
  {\bibfield  {journal} {\bibinfo  {journal} {Phys. Rev. Lett.}\ }\textbf
  {\bibinfo {volume} {92}},\ \bibinfo {pages} {035301} (\bibinfo {year}
  {2004})}\BibitemShut {NoStop}%
\bibitem [{\citenamefont {Laurie}\ \emph {et~al.}(2012)\citenamefont {Laurie},
  \citenamefont {Bortolozzo}, \citenamefont {Nazarenko},\ and\ \citenamefont
  {Residori}}]{LBNR2012}%
  \BibitemOpen
  \bibfield  {author} {\bibinfo {author} {\bibfnamefont {J.}~\bibnamefont
  {Laurie}}, \bibinfo {author} {\bibfnamefont {U.}~\bibnamefont {Bortolozzo}},
  \bibinfo {author} {\bibfnamefont {S.}~\bibnamefont {Nazarenko}},\ and\
  \bibinfo {author} {\bibfnamefont {S.}~\bibnamefont {Residori}},\ }\href
  {https://doi.org/https://doi.org/10.1016/j.physrep.2012.01.004} {\bibfield
  {journal} {\bibinfo  {journal} {Physics Reports}\ }\textbf {\bibinfo {volume}
  {514}},\ \bibinfo {pages} {121} (\bibinfo {year} {2012})},\ \bibinfo {note}
  {one-Dimensional Optical Wave Turbulence: Experiment and Theory}\BibitemShut
  {NoStop}%
\bibitem [{Note1()}]{Note1}%
  \BibitemOpen
  \bibinfo {note} {The leading-order terms in the small-amplitude expansion of
  the governing equations in both Refs.~\cite {KS2004} and~\cite {LBNR2012}
  yield the dispersion relation $\omega =k^2$ in an appropriate scale range.
  For this relation, it is known that no resonant interactions among groups of
  four waves exist in one spatial dimension. Therefore, the cubic terms, which
  are the first nonlinear correction in the governing equation, can be
  eliminated using a canonical transformation~\cite {ZLF1992}. The next-order,
  quintic terms, together with the corresponding resonant interactions among
  groups of six waves, thus dominate the dynamics. Resonant interactions among
  groups of eight waves stem from yet higher-order terms and can thus be
  neglected.}\BibitemShut {Stop}%
\bibitem [{\citenamefont {Buckmaster}\ \emph {et~al.}(2018)\citenamefont
  {Buckmaster}, \citenamefont {Germain}, \citenamefont {Hani},\ and\
  \citenamefont {Shatah}}]{BuGeHaSh18}%
  \BibitemOpen
  \bibfield  {author} {\bibinfo {author} {\bibfnamefont {T.}~\bibnamefont
  {Buckmaster}}, \bibinfo {author} {\bibfnamefont {P.}~\bibnamefont {Germain}},
  \bibinfo {author} {\bibfnamefont {Z.}~\bibnamefont {Hani}},\ and\ \bibinfo
  {author} {\bibfnamefont {J.}~\bibnamefont {Shatah}},\ }\href
  {https://doi.org/https://doi.org/10.1002/cpa.21749} {\bibfield  {journal}
  {\bibinfo  {journal} {Communications on Pure and Applied Mathematics}\
  }\textbf {\bibinfo {volume} {71}},\ \bibinfo {pages} {1407} (\bibinfo {year}
  {2018})},\ \Eprint
  {https://arxiv.org/abs/https://onlinelibrary.wiley.com/doi/pdf/10.1002/cpa.21749}
  {https://onlinelibrary.wiley.com/doi/pdf/10.1002/cpa.21749} \BibitemShut
  {NoStop}%
\bibitem [{Note2()}]{Note2}%
  \BibitemOpen
  \bibinfo {note} {We note that with $-1<p<0$ fixed, the requirement $\tau
  _{kin}\gg 1$ provides an additional practical bound on $L$ for our
  simulations, which degenerates at $p=0$.}\BibitemShut {Stop}%
\bibitem [{\citenamefont {Webb}(1978)}]{Webb1978}%
  \BibitemOpen
  \bibfield  {author} {\bibinfo {author} {\bibfnamefont {D.}~\bibnamefont
  {Webb}},\ }\href {https://doi.org/10.1016/0146-6291(78)90593-3} {\bibfield
  {journal} {\bibinfo  {journal} {Deep Sea Research}\ }\textbf {\bibinfo
  {volume} {25}},\ \bibinfo {pages} {279} (\bibinfo {year} {1978})}\BibitemShut
  {NoStop}%
\bibitem [{\citenamefont {Tracy}\ and\ \citenamefont {Resio}(1982)}]{TR1982}%
  \BibitemOpen
  \bibfield  {author} {\bibinfo {author} {\bibfnamefont {B.}~\bibnamefont
  {Tracy}}\ and\ \bibinfo {author} {\bibfnamefont {D.}~\bibnamefont {Resio}},\
  }\href@noop {} {\bibfield  {journal} {\bibinfo  {journal} {US Army Corps of
  Engineers: Washington, DC, USA}\ }\textbf {\bibinfo {volume} {{WIS} Report
  11}} (\bibinfo {year} {1982})}\BibitemShut {NoStop}%
\bibitem [{\citenamefont {Resio}\ and\ \citenamefont {Perrie}(1991)}]{RP1991}%
  \BibitemOpen
  \bibfield  {author} {\bibinfo {author} {\bibfnamefont {D.}~\bibnamefont
  {Resio}}\ and\ \bibinfo {author} {\bibfnamefont {W.}~\bibnamefont {Perrie}},\
  }\href@noop {} {\bibfield  {journal} {\bibinfo  {journal} {J. Fluid Mech.}\
  }\textbf {\bibinfo {volume} {223}},\ \bibinfo {pages} {603} (\bibinfo {year}
  {1991})}\BibitemShut {NoStop}%
\end{thebibliography}

%
\end{document}